\documentclass[12pt,a4paper]{article}
\usepackage{amsmath}
\input epsf
\usepackage[dvips]{graphicx}
\usepackage{times}

\begin{document}
\begin{titlepage}
\title{Unitarity and the color confinement}
\author{ S.M. Troshin,
  N.E. Tyurin\\[1ex] \small\it Institute
\small\it for High Energy Physics,\\\small\it Protvino, Moscow Region, 142281 Russia}
\date{}
\maketitle
\begin{abstract}
We discuss  how confinement property of QCD results in the rational
 unitarization scheme and how unitarity saturation leads to appearance of a hadron liquid phase at very high temperatures. 
\end{abstract}
\vfill
\end{titlepage}

\section*{Introduction}
One of  the fundamental
problems of  QCD  is related to a confinement  of the color. This phenomenon
is associated with collective, coherent interactions of quarks and gluons,
and results in formation of the asymptotic states, which are the
colourless, experimentally observable particles. 

On the other, theoretical side, there is a hypothesis 
on the completeness of a set of the asymptotic hadronic states. This assumption plays an important role  (cf. \cite{bart})
 in strong interaction  theory
and leads, e.g. due to  unitarity of the scattering matrix, to an optical theorem relating the total cross-section
with  the  value of the elastic scattering amplitude in the forward direction, i.e. at $t=0$.
The unitarity condition is formulated
for  the asymptotic colorless  on-mass shell states and does not  constrain
the  fundamental fields of QCD -- colored fields of quarks and gluons.

The Hilbert space, in the axiomatic formulation of the quantum field theory,
 corresponds to the colorless hadron states. It is constructed using vectors obtained by 
action  of the relevant creation operators on the physical vacuum. The state of  physical vacuum 
is the state without particles, i.e. annihilation operator acting on it produces zero. It is  
the state of a lowest energy of the system.
This state is also invariant under Lorentz transformations. 

The  point of completeness of a set of the asymptotic states, which includes hadrons only, can be considered as a 
questionable one in the QCD, and the unitarity could be violated in the above indicated    sence (cf. \cite{rlst}).
It was stated that the Hilbert space, which corresponds to colorless hadron states and is constructed using vectors
  spanned
on the physical vacuum state, should be extended. 
Indeed, nowadays it is  often regarded that the vacuum state  is not
a unique one: colored  current quarks and gluons
are the degrees of freedom related to  another,  perturbative vacuum. The vacuum state is not considered anymore as a state
of the lowest energy of a system. 

According to the confinement property 
of QCD, isolated colored objects cannot exist in the physical vacuum and  there is no room for the objects like quark-proton scattering amplitude, 
since isolated color object has an infinite energy in the physical vacuum.
  Transition from physical vacuum 
to the perturbative one supposed to occur in the  process of deconfinement resulting in quark-gluon plasma
formation, i.e.   a gaseous state of the free colored quarks and gluons.  Thus, it is evident that
the hadrons  and free quarks (and gluons)  cannot coexist  together since
those  belong to the  different vacua \cite{mingmei}.

In this paper we address the issues related to unitarity and its relation to the confinement property of QCD. In the next section we demonstrate
 how the color confinement could result in the rational unitarizarion scheme for the scattering matrix. 
Section 2 is devoted to discussion of the consequences of the rational unitarization scheme, namely appearance of 
the reflective scattering mode. It will be shown that this mode could result in the emerging new phase of the hadron matter --- hadron liquid. 
In the section 3 a possible microscopic mechanism for the above transition and existence of the third vacuum state is
 discussed in the context of the unitarity saturation at very high energies.
\section{Unitarity and confinement}
In this section we  demonstrate that inclusion into a set of the asymptotic states of the fields corresponding to the confined 
objects would lead to a rational form of S-matrix unitarization provided those fields satisfy a certain simple constraint. 
We address  above issues using a paper by N.N. Bogolyubov \cite{nnb}\footnote{The reference 
to this paper
was brought to our attention by V.A. Petrov.}  as a guide and consider a state vector $|\Phi\rangle$ being a sum of the two vectors
\[
 |\Phi \rangle=|\varphi\rangle+|\omega\rangle.
\]
where $|\varphi\rangle$ corresponds to the physical states and belongs to the Hilbert space ${\cal H}_\varphi$ and $|\omega\rangle$ -- 
to the confined states and
belongs to the Hilbert space ${\cal H}_\omega$.  So, we  have that $|\varphi\rangle={\cal P}|\Phi\rangle$ 
and $|\omega\rangle=(1-{\cal P})|\Phi\rangle$, 
where $\cal P$ is a relevant projection operator.
 Difference with consideration performed in \cite{nnb} is in the replacement of the
states with indefinite metrics  by the  states of confined objects such as quarks and gluons.
A common Hilbert space $\cal H$ is a sum:
\[
\cal H={\cal H}_\varphi+{\cal H}_\omega
\]
and the scattering operator $\cal \tilde S$  acting in $\cal H$,
\[
|\Phi,{out}\rangle={\cal \tilde S}|\Phi,{in}\rangle,
\]
should not, in priciple, conserve probability and obey unitarity condition
since $\cal H$ includes ${\cal H}_\omega$ -- Hilbert space where confined objects reside.  
The norm of  confined objects  is not defined.  

Next, let us to impose condition similar to the one used in \cite{nnb}, i.e.
\[
| \omega,{in}\rangle+|\omega,{out}\rangle=0.
\]
 It means that {\it in-} and {\it out-} vectors corresponding to the states of the confined objects are just the mirror reflections of each other.
 Those reflections can be associated with the impossibility for confined objects to propogate outside the hadron border.
Thus, the rational form of unitary scattering operator $\cal S$ 
\[
 {\cal S}=(1-{\cal U})(1+{\cal U})^{-1},
\]
in the physical Hilbert space ${\cal H}_\varphi$,
\[ 
|\varphi,{out}\rangle={\cal S}|\varphi,{in}\rangle
\]
can easily be obtained, since
\[
|\varphi,{out}\rangle={\cal P}{\cal \tilde S}( |\varphi,{in}\rangle+ |\omega,{in}\rangle).
\]
Operator $\cal U$ has the following form 
\[
{\cal U}= (1-{\cal P}){\cal \tilde S}.
\]

 We started with non-unitary scattering operator ${\cal \tilde S}$ and obtained
the scattering operator ${\cal  S}$, which automatically satisfy unitarity condition.
Crucial assumption there was a constraint for the states of confined objects $| \omega,{in}\rangle+|\omega,{out}\rangle=0$, which
we assumed to be  a cofinement condition since it provides a condition for an existence of a horizon for the colored particles. 
Colorless particles such as a photon or a pion are not affected by the existence of such a horizon and can travel beyond hadron boundaries. 
Thus, it is very tempting to claim  that unitarity can be related to a confinement.

Rational or $U$-matrix form of unitarization was proposed long time ago \cite{heit} in the theory of radiation dumping.
Self-damping of inelastic channels was considered in \cite{bbla} and
 for the  relatvistic case such form of unitarization was obtained in \cite{umat}. 
But, an importance of the forgotten paper \cite{nnb} is, in particular, the following: 
it provides a clue for the physical interpretation of $U$-matrix. Nowadays it can be used for construction of a bridge
between the physical states of hadrons and the states of confined objects -- quarks and gluons. 

The elastic scattering $S$-matrix, i.e. the $2\to 2$  matrix element of the operator $\cal S$,
in the impact parameter representation can be
written in this  unitarisation scheme in the form of linear 
rational transform (cf. \cite{inta} and references therein) and in
the case of pure imaginary $U$-matrix
\begin{equation}
S(s,b)={[1-U(s,b)]}/{[1+U(s,b)]}, \label{um}
\end{equation}
where $U(s,b)$ is the generalized reaction matrix. 
It is
considered to be an input dynamical quantity. And, this is an essential point, an explicit form of the function $U(s,b)$ and the numerical
predictions for the observable quantities depend on the particular  model used for hadron scattering description.

With account of what was said above we can associate this function with matrix elements\footnote{Imaginary part of $U(s,b)$ 
gets contributions from inelastic intermidiate channels.}
 of the operator
$(1-{\cal P}){\cal \tilde S}$, i.e. $U(s,b)$ should be related to a scattering dynamics of the confined hadron constituents.

Rational representation of the scattering matrix leads to several distinctive features such as  a peripherality of inelastic processes \cite{inta} 
and restoration of confined phase of hadronic
matter at very high temperatures.  We consider the latter issue in the following section.

\section{Hadronic liquid at very high temperatures}
For the qualitative purposes it
is sufficient to know \cite{inta} that the function $U(s,b)$ increases with energy in a power-like way
 and decreases with impact parameter like a linear exponent or Gaussian \footnote{In fact, the analytical properties
of the scattering amplitude imply a linear exponential dependence at large values of $b$.}.
It can  easily be seen that the new scattering mode, reflective scattering (when $S(s,b)<0$) starts to appear at
the energy $s_R$, which is determined by a solution of the equation
\[
U(s_R,b=0)=1.
\]
Indeed, the unitarity relation
written for the elastic scattering amplitude $f(s,b)$  in the high
energy limit has the following form
\begin{equation}
\mbox{Im} f(s,b)=h_{el}(s,b)+h_{inel}(s,b). \label{ub}
\end{equation}
The inelastic overlap function $h_{inel}(s,b)$
is connected with the function $U(s,b)$ by the relation
\begin{equation}\label{hiu}
h_{inel}(s,b)={ U(s,b)}/{[1+U(s,b)]^{2}},
\end{equation}
and the only condition to obey unitarity
 is $ U(s,b)\geq 0$. 
The elastic overlap function in its turn is related to the function
 $U(s,b)$ as follows
\begin{equation}\label{heu}
h_{el}(s,b)={[U(s,b)]^{2}}/{[1+U(s,b)]^{2}}.
\end{equation}

At sufficiently high energies the inelastic overlap function
 $h_{inel}(s,b)$ would have a peripheral $b$-dependence and will
tend to zero for $b=0$ at $s\to\infty$ (cf. e.g \cite{inta}).
Therefore, corresponding behaviour of the elastic scattering
$S$-matrix (note that $S(s,b)=1+2if(s,b)$) can then be interpreted
as an appearance of a reflecting ability of scatterer due to increase of
 its density beyond some critical value.  In another words, the scatterer has now not only
 absorption ability (due to  presence of inelastic channels), but it starts to be reflective at very
 high energies. In central collisions, $b=0$, an elastic scattering
  approaches to the completely
 reflecting limit $S=-1$ at $s\to\infty$. 

At the  energy values $s>s_R$ the equation $U(s,b)=1$ has a solution in the
physical region of the impact parameter values, i.e. $S(s,b)=0$ at $b=R(s)$. 
The probability of reflective scattering at $b<R(s)$ and $s> s_R$ is determined by the magnitude
 of $|S(s,b)|^2$; this probability is equal to zero at $s\leq s_R$ and $b\geq R(s)$. Those inequalities impose an equation for
a horizon for the reflective scattering events.
The dependence of $R(s)$
 is determined  by the logarithmic function,  $R(s) \sim \frac{1}{M}\ln s$ . This dependence
 is consistent with analytical properties of the resulting elastics scattering amplitude in
  the complex $t$-plane and mass $M$ can be related to the pion mass.
Thus, at the energies $s> s_R$ the reflective scattering will simulate presence of the repulsive core in
the hadron and meson interactions and the reflective elastic scattering will become a dominating process
at sufficiently high energies. 
This kind of the elastic scattering preserving
the hadron identities  acts against deconfinement. 
It would lead to a new phase of the hadron matter at very high temperatures. 

Presence of the reflective scattering can be
 accounted for by using the van der Waals method (cf. \cite{cleym}).
 This approach originally was used   for the  description of  liquids  starting from
 a gas approximation by introducing a nonzero size of molecules. 

The hadron liquid density $n_R(T,\mu)$  can be connected then \cite{limdens}
with the density in the  approach without reflective scattering (i.e. poinlike type of interaction) $n(T,\mu)$
 by the following relation
\[
  n_R(T,\mu)={n(T,\mu)}/{[1+\kappa(s)n(T,\mu)]},
\]
where $\kappa(s)={p_R(s)V_R(s)}/{2}$, 
$p_R(s)$ is the averaged over volume $V_R(s)$ probability of reflective scattering
and the volume $V_R(s)$ is determined by the radius of the reflective scattering.
At very high energies ($s\to\infty$)
\[
n_R(T,\mu)\sim 1/\kappa(s)\sim M^3/\ln ^3 s.
\]
This  limiting dependence for the hadron liquid density  appears due to 
presence of the reflective scattering. In the oversimplified geometrical picture it resembles a
scattering of hard spheres in the head-on hadron collisions .
 It can also be associated with  saturation of the Froissart-Martin bound for the total cross-section. 

At very high temperatures we could expect that a new confined phase corresponding to hadron
liquid reappears. A possible presence of this phase is connected with the unitarity  saturation, and
a relevant microscopic mechanism will be discussed in the next section.

\section{Microscopic mechanism}
The picture described above implies an existence of the two vacuum states: perturbative and physical ones. It results
from assumption on the same scale of transitions responsible for confinement-deconfinement and chiral restoration. 
This assumption has a theoretical ground in some of the lattice calculations (cf. e.g. \cite{born}).  

However, it is often assumed
that the scales relevant to confinement and chiral symmetry breaking are different \cite{manoh}, scale of confinement 
$\Lambda_{QCD} = 100-300$ MeV while chiral symmetry breaking scale --- $\Lambda_\chi\simeq 1$ GeV. 
Thus, in the range 
between these two scales the matter is in a deconfined state but chiral symmetry is spontaneously broken there. 
In  line with this picture, which can be treated as a posteriori justification, long time ago, in the pre-QCD era, 
it was supposed that hadrons have a simple structure and
non-relativistic quark model has been commonly adopted. During recent time  such a model has evolved and obtained much more
solid theoretical grounds \cite{manoh,kaplan,diak}. As it will be discussed further, one can assume existence in the hadron's interior
of the third (nonperturbative) vacuum state with colored constituent quarks
and pions as the relevant degrees of freedom.

The origin of a nonperturbative vacuum and relevant effective degrees of freedom can be related to
the mechanism of spontaneous chiral symmetry breaking ($\chi$SB) in QCD \cite{bjorken}.
  This mechanism describes
transition of the current into  the constituent quarks and emerging of the Goldstone bosons.
Massive  constituent quarks appear  as quasi-particles, i.e. current quarks and
the surrounding  clouds  of quark--antiquark pairs.  
The constituent quarks interact  via exchange
of the Goldstone bosons; this interaction is mainly due to a pion field.
Pions themselves are the bound states of massive
quarks.
Constituent quark interaction with the Goldstone bosons  is strong and could have 
  the following form  \cite{diak}:
 \begin{equation}
{\cal{L}}_I=\bar Q[i\partial\hspace{-2.5mm}/-M\exp(i\gamma_5\pi^A\lambda^A/F_\pi)]Q,\quad \pi^A=\pi,K,\eta.
\end{equation}
For simplicity, in what follows we refer to  pions only, denoting by this 
generic word all Goldstone bosons, i.e. pions themselves, kaons
 and $\eta$-mesons.

Thus, we will assume that  the vacuum state $|0\rangle_{pt}$ has a perturbative nature at short distances
with current quarks and gluons as degrees of freedom, at large distances the physical vacuum state 
$|0\rangle_{ph}$ has relevant colorless hadrons as degrees of freedom, and inside a hadron  the vacuum $|0\rangle_{np}$
has a nonperturbative origin with constituent quarks and Goldstone pions being the relevant degrees of freedom.
We suppose the picture of a hadron consisting of constituent quarks  interacting with pions. The latter 
have a  dual role: Goldstone and physical particles. 

There are different approaches to the deconfinement dynamics.   It was recently   proposed \cite{satz,satz2} to use
percolation theory as a candidate for the mechanism of deconfinement in the form of the analytical
crossover (i.e. without first and second order phase transitions).  This form of deconfinement 
was found in the experimental studies at RHIC.
 Evidently, such  purely geometrical perolation approach should be amended by  a 
dynamical mechanism and
color dynamics of deconfinement due to formation
of molecular-like aggregations was proposed in \cite{mingmei}. The vacuum 
inside the hadron was taken to be a perturbative one and quark interactions have
origin in the colour dynamics. It seems, however, that for crossover nature of deconfinement dynamics 
 it is more natural to
expect transition of physical to nonperturbative vacuum $|0\rangle_{ph}\to |0\rangle_{np}$ instead of transition $|0\rangle_{ph}\to |0\rangle_{pt}$. 
Indeed, using effective quark-pion interaction inside hadron and hadron-pion interaction outside
hadron, we have a pion field as an universal interaction agent for both confined and deconfined states
 and this could serve as a natural explanation of deconfinement as a crossover transition.

Experimentally, deconfined state of matter has been discovered  
at RHIC where the highest values of energy and
density have been reached. This deconfined
state appears to be a strongly interacting collective state
with properties of a perfect liquid.   
 The matter is strongly
correlated and reveals the high degree of  coherence when it is
 well beyond the critical values of density and
temperature. In the framework of the approach under consideration this state
can be interpreted as a Quark (constituent)-pion liquid in the nonperturbative
vacuum $|0\rangle_{np}$. 

A natural question arises then: what one should expect at higher temperatures, e.g. at the LHC energies, i.e.
would one observe transition $|0\rangle_{np}\to |0\rangle_{pt}$, or additional possibilities exist? 
Indeed, due to a large 
kinetic energy of the constituent quarks in the nonperturbative vacuum
there should be a finite probability to form the colorless clusters again,
i.e. confinement mechanism could take place, transition $|0\rangle_{np}\to |0\rangle_{ph}$
 would happen instead of the transition   $|0\rangle_{np}\to |0\rangle_{pt}$  and hadrons would reappear
again. 
This  possibility
obtain support from unitarity saturation at very high energies discussed in the previous section.

Thus, the following chain of  phase transitions of hadron matter could exist as the temperature increases at the constant value of chemical
potential $\mu$:
\[
 |0\rangle_{ph}(\mbox{Hadron gas})\to |0\rangle_{np}(\mbox{Quark-pion liquid})\to |0\rangle_{ph}(\mbox{Hadron liquid}).
\]
In the above chain  a physical vacuum suffers a loop transition to a nonperturbative one and back to a physical vacuum.
It is also useful to keep in mind that due to Lorentz invariance of the vacuum states energy difference between those 
states should also be Lorentz invariant \cite{bart}, i.e. several vacua would be either degenerate or have infinite energy difference.  
On that basis it can be supposed that crossover form of deconfinement implies degeneracy of the states: 
$|0\rangle_{ph}$ and $|0\rangle_{np}$.

The picture decribed above is certainly an oversimplified one. The reflective scattering is always accompanied 
by the absorbtive scattering at moderate and large impact parameters in hadronic collisons. Such highly energetic peripheral 
hadron collisions might imply
the transitions $|0\rangle_{np}\to |0\rangle_{pt}$ and, in fact, at very high temperatures we might have not a homogeneous phase of 
a hadronic liquid.

\section*{Conclusion}
It was shown how unitary form of the scattering matrix could be inhereted from the confinement property of QCD.  
It was conjectured also that  saturation of unitarity  would restore confinement, and
percolation mechanism alone is not sufficient for the deconfinement as it was assumed in \cite{satz,satz2,cleym,limdens}. In general,
we would like to note that at very high temperatures there is a certain probability that the matter would return
to a confined state if the unitarity saturation would occur there. 
Hopefully, the experimental studies with heavy ions at the LHC will be able to reveal the new  phases of hdronic matter.

\section*{Acknowledgement}
We gratefully acknowledge stimulating discussions and correspondence with\\
L.L.~Jenkovszky, K.G.~Klimenko, V.A.~Petrov, and J.P.~Ralston.

\end{document}